\documentclass{elsart5p}
 \usepackage{graphicx}

\usepackage{amssymb}

\journal{J. Magn. Magn. Mater.}

\begin{document}

\begin{frontmatter}


 \title{Monte Carlo investigation of the magnetic anisotropy in Fe/Dy multilayers}

 \author{E. Talbot\corauthref{cor1}}
 \ead{etienne.talbot@univ-rouen.fr}
 \corauth[cor1]{Tel: + 33-2-32-95-51-32    ; fax: + 33-2-32-95-50-32}
 \author{P.E. Berche}
 \author{D. Ledue}

 \address{Groupe de Physique des Mat\'{e}riaux, Universit\'{e} de Rouen, UMR 6634 CNRS; avenue de l'universit\'{e} - BP12 76801 Saint Etienne du Rouvray, France}

\begin{abstract}
By Monte Carlo simulations in the canonical ensemble, we have studied the magnetic anisotropy in Fe/Dy amorphous multilayers. This work  has been motivated by experimental results which show a clear correlation between the magnetic perpendicular anisotropy and the substrate temperature during elaboration of the samples. Our aim is to relate macroscopic magnetic properties of the multilayers to their structure, more precisely their concentration profile. Our model is based on concentration dependent exchange interactions and spin values, on random magnetic anisotropy and on the existence of locally ordered clusters that leads to a perpendicular magnetisation. Our results evidence that a compensation point occurs in the case of an abrupt concentration profile. Moreover, an increase of the noncollinearity of the atomic moments has been evidenced when the Dy anisotropy constant value grows. We have also shown the existence of inhomogeneous magnetisation profiles along the samples which are related to the concentration profiles.
\end{abstract}

\end{frontmatter}

\section{Introduction}
\label{intro}
Amorphous multilayers of transition metal - rare earth (TM-RE) compounds display particular interesting properties such as a strong magnetic anisotropy which, in some conditions, may be perpendicular to the layers \cite{sato86,shan90}. Up to now, the accurate origin of this anisotropy is not clearly understood and several models have been proposed. These models are based on the anisotropic distribution of the TM-RE pairs \cite{sato86}, dipolar interactions \cite{schulte95}, local structural anisotropy \cite{mergel93,fuji96} or single-ion anisotropy \cite{shan90b}. However, no model is able to explain all the experimental features.

\begin{figure}
\centering
\rotatebox{-90}{\includegraphics[width=0.33\textwidth]{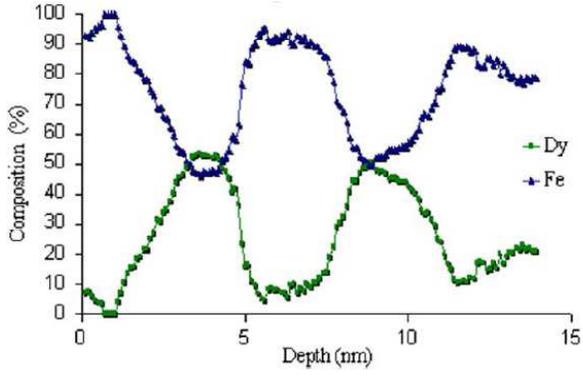}}
\caption{Concentration profile of an (Fe 3nm/ Dy 2nm) multilayer deposited at 570K. The data have been obtained by tomographic atom probe measurements \cite{tamion06}.}
\label{figure01}
\end{figure}

This study has been performed by taking into account recent experimental results of the structural and of the magnetic properties of amorphous Fe/Dy multilayers \cite{tamion05,tamion06,tamion07}. In particular, hysteresis loops measured with the magnetic field applied successively in the film plane and perpendicular to it show that the perpendicular magnetic anisotropy is stronger for the multilayers deposited at a higher temperature (600K compared to 320K) \cite{tamion06}. In this temperature range, the structural investigations have shown that the multilayer is alternatively composed of layers with a majority of amorphous Fe and of Fe-Dy alloyed layers (Figure \ref{figure01}) \cite{tamion06}. For $T_{S}=320$K, the multilayer is made up of roughly pure amorphous Fe and Dy layers with thin interfaces. Above 700K, the layered structure is destroyed, replaced by mixed grains of $\alpha$-Fe and Fe$_2$-Dy which leads to a planar anisotropy. Altogether, these results show that the perpendicular magnetic anisotropy seems to be correlated in these samples with the existence of Fe-Dy amorphous alloy in the multilayers.

In this work, we have performed Monte Carlo (MC) simulations on amorphous Fe/Dy multilayers in order to investigate the influence of the concentration profile on the magnetic properties. We have considered a model including perpendicular magnetic aniso\-tropy (on average) on a small fraction of Dy sites, the other Dy sites exhibit random anisotropy. Two different concentration profiles have been studied: an abrupt profile that is the multilayer is made up of pure Fe layers and pure Dy layers (close to the experimental profile for $T_{S}=320$K) and an experimental profile obtained from tomographic atom probe analysis on a multilayer deposited at 570K (Figure \ref{figure01}). Our goal is to investigate by means of MC simulations the influence of the concentration profile on the global magnetisation, on the global susceptibility, on the sublattice magnetisations and on the magnetisation per plane. Since the strength of the Dy single site anisotropy is not clearly elucidated, we have also investigated the influence of the Dy anisotropy constant magnitude. Moreover, the fraction of Dy sites with perpendicular magnetic anisotropy is not known, so we have also let vary this parameter in our simulations to determine its influence. The model is presented in section \ref{sec2} and the simulation method is described in section \ref{sec3}. The results are discussed in section \ref{sec4} and a conclusion is given in section \ref{sec5}.

\section{Model} \label{sec2}
\subsection{Magnetic parameters}
\begin{figure}[ht]\centering
\rotatebox{-90}{\includegraphics[width=0.3\textwidth]{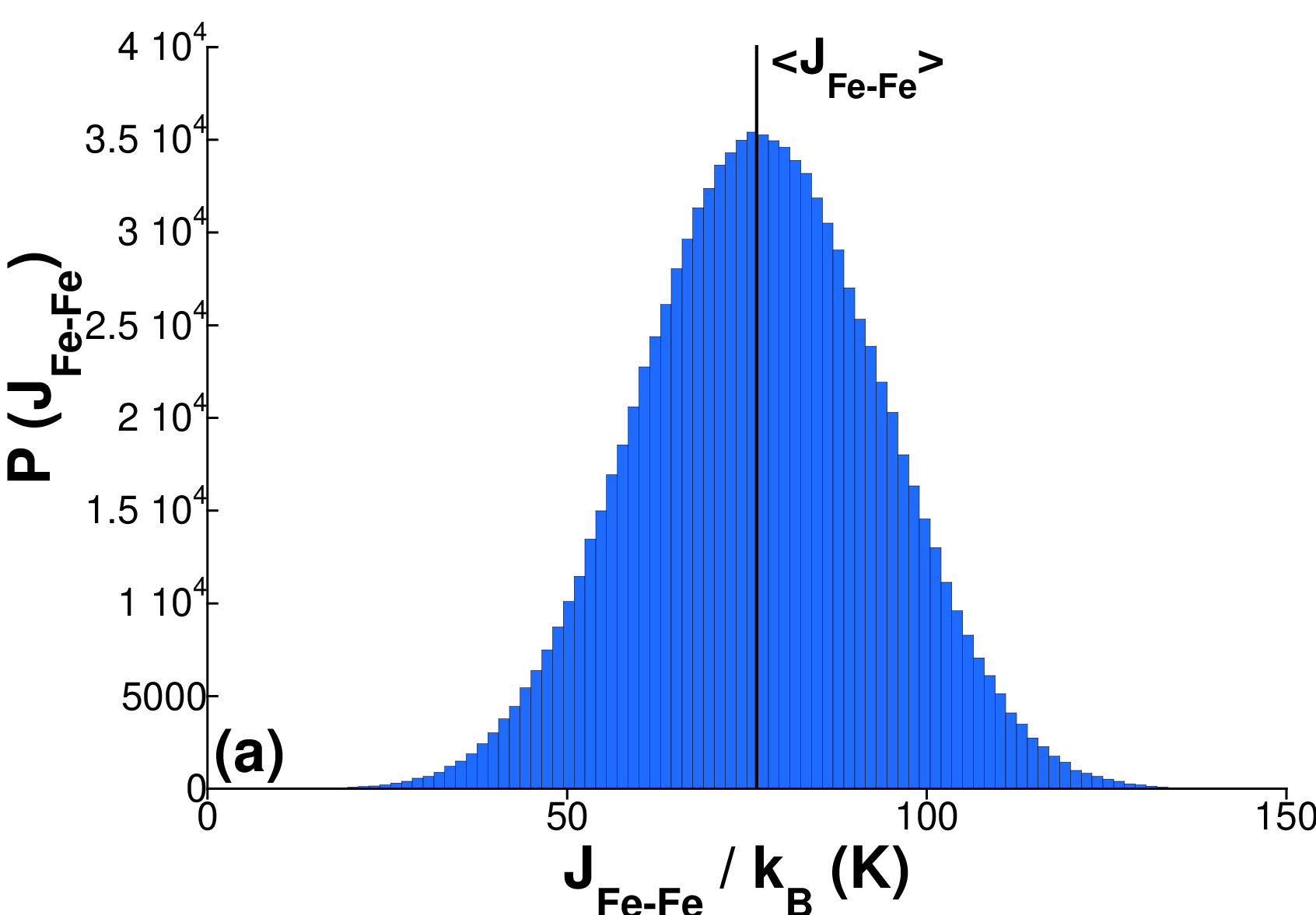}}
\rotatebox{-90}{\includegraphics[width=0.3\textwidth]{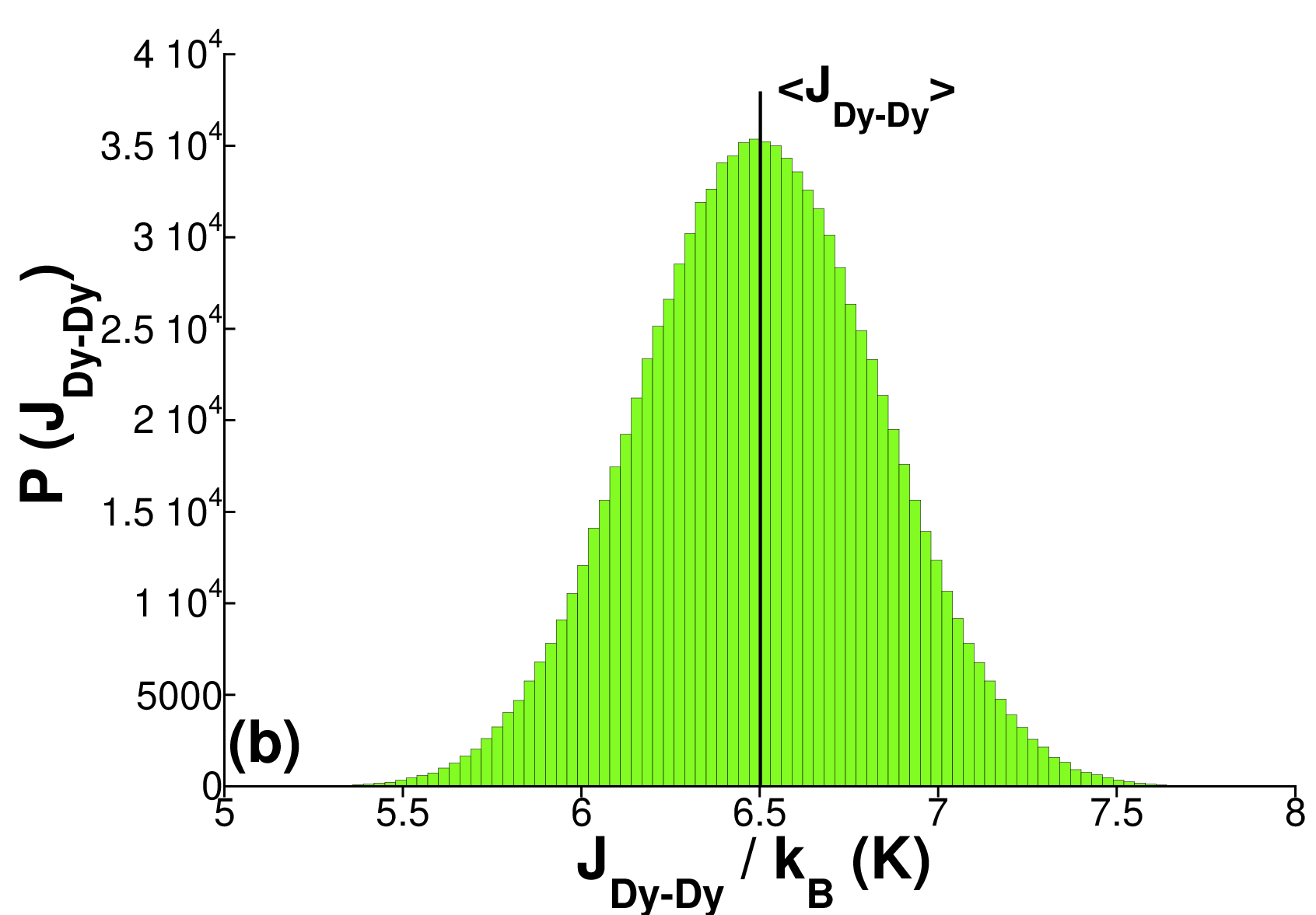}}
	\caption{Gaussian distribution of the exchange interaction ((a) Fe-Fe exchange interaction; (b) Dy-Dy exchange interaction).}
\label{figure02}
\end{figure}

Our model of Fe/Dy ferrimagnetic multilayers consists of classical Heisenberg spins located at the sites of a face centered cubic lattice with nearest-neighbour exchange interactions. We have chosen this lattice in order to reproduce the experimental observed compact structure \cite{heiman76,mimura78,hansen91}. As the multilayers are amorphous, the exchange interactions are modulated by a gaussian distribution (Figure \ref{figure02}) which models the distribution of the interatomic distances in real samples \cite{kobe76}. On the other hand, the $J_{{\rm Fe-Fe}}$ and $J_{{\rm Fe-Dy}}$ exchange interaction values have been extracted from those determined by Heiman et al. \cite{heiman76} on Fe-Dy amorphous thin films by mean-field calculations and adjusted  by MC simulations in order to get the pure amorphous Fe Curie temperature (270K \cite{heiman76}). So, they linearly depend on the local concentration $X_{{\rm Fe}}$ following:

$$J_{{\rm Fe-Fe}}(X_{{\rm Fe}})/k_{\rm B}=77+449 (1-X_{{\rm Fe}})\ {\rm (in\ K)} \quad (X_{\mathrm{Fe}}>0.4),$$
$$J_{{\rm Fe-Dy}}(X_{{\rm Fe}})/k_{\rm B}=8-198 (1-X_{{\rm Fe}})\ {\rm (in\ K)} \quad (X_{\mathrm{Fe}}>0.4).$$
Concerning the $J_{{\rm Dy-Dy}}$ exchange interaction which is known to be much smaller than the others, this latter has been taken constant in function of the concentration. Its value has been estimated by MC simulations in order to provide the pure amorphous Dy Curie temperature (110K \cite{tappert96}):
$$J_{{\rm Dy-Dy}}/k_{\rm B}=6.5\ {\rm (K)}.$$
These concentration rules are valid only when $X_{\mathrm{Fe}}>0.4$. $J_{{\rm Fe-Dy}}$ is strongly negative; it is thus responsible for the ferrimagnetic order (with antiparallel Fe and Dy moments) at low temperature which has been experimentally observed by polarized neutrons reflectivity measurements \cite{tamion07}. We have to mention that the Fe spin value also depends on the local concentration  \cite{heiman76}:
$$S_{{\rm Fe}}(X_{{\rm Fe}})=1.1-1.125(1-X_{{\rm Fe}}) \quad (X_{\mathrm{Fe}}>0.4).$$
The other numerical values of the model are those of the free ions:
$$g_{{\rm Fe}}=2;\ g_{{\rm Dy}}=4/3;\ S_{{\rm Dy}}=2.5.$$

\begin{figure}[ht]\centering
\rotatebox{-90}{\includegraphics[width=0.3\textwidth]{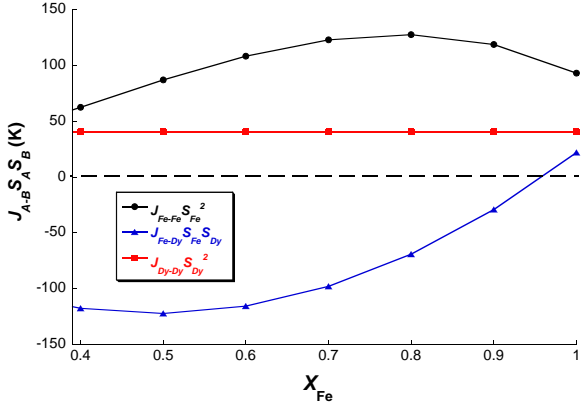}}
	\caption{Exchange energy as a function of the Fe concentration for the three bonds.}
\label{figure03}
\end{figure}

The variation of the exchange energy as a function of the Fe concentration for the three different bonds is shown in Figure \ref{figure03}. We can see that the exchange energy (in absolute value) of the Fe-Fe and Fe-Dy bonds are maximum respectively at $X_{\mathrm{Fe}}=0.8$ and $X_{\mathrm{Fe}}=0.5$ and that the contribution of the Fe-Dy pairs will play a significant role in the magnetic ordering.

\subsection{Concentration profile}
\begin{figure}[ht]\centering
\rotatebox{-90}{\includegraphics[width=0.3\textwidth]{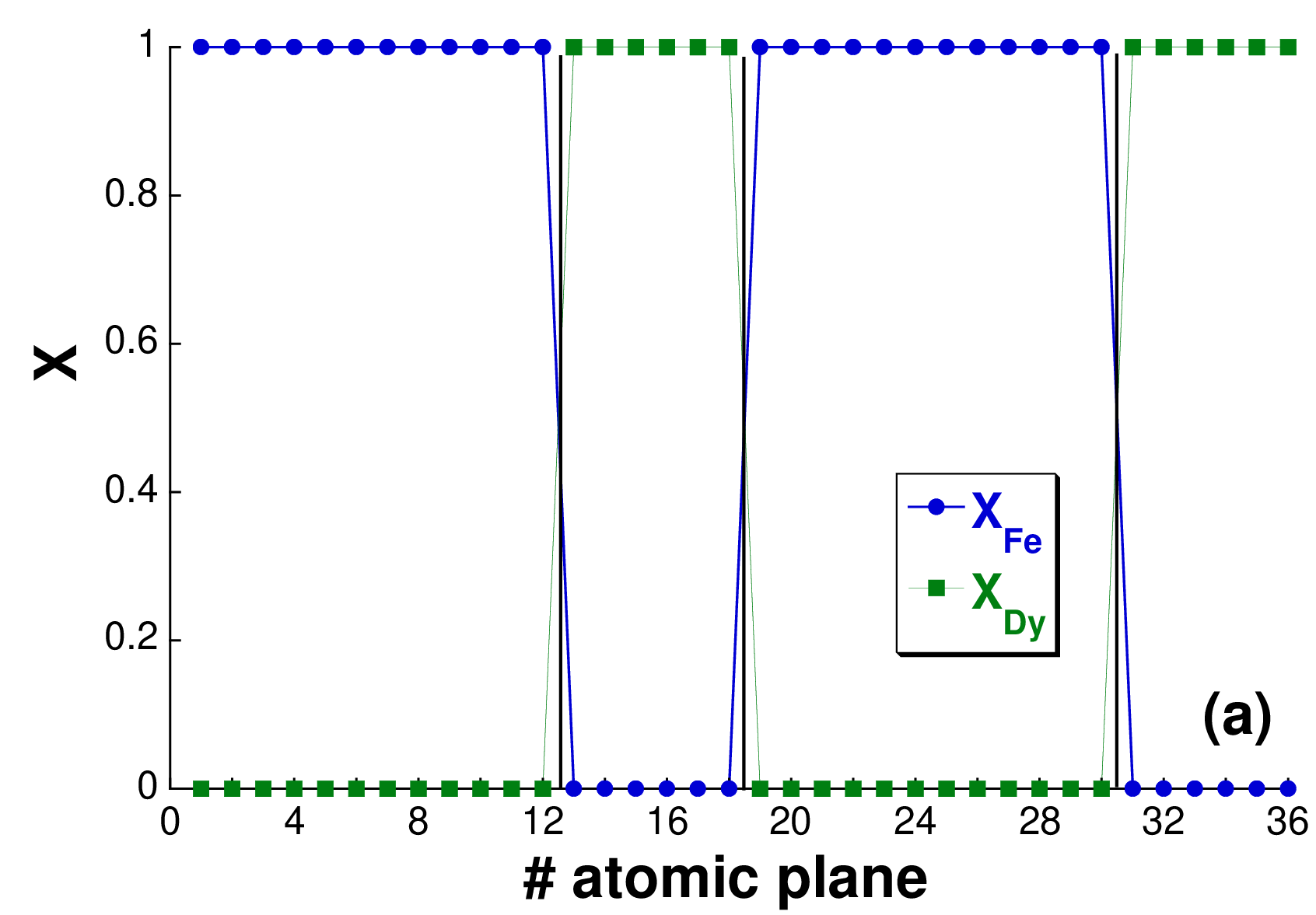}}
\rotatebox{-90}{\includegraphics[width=0.3\textwidth]{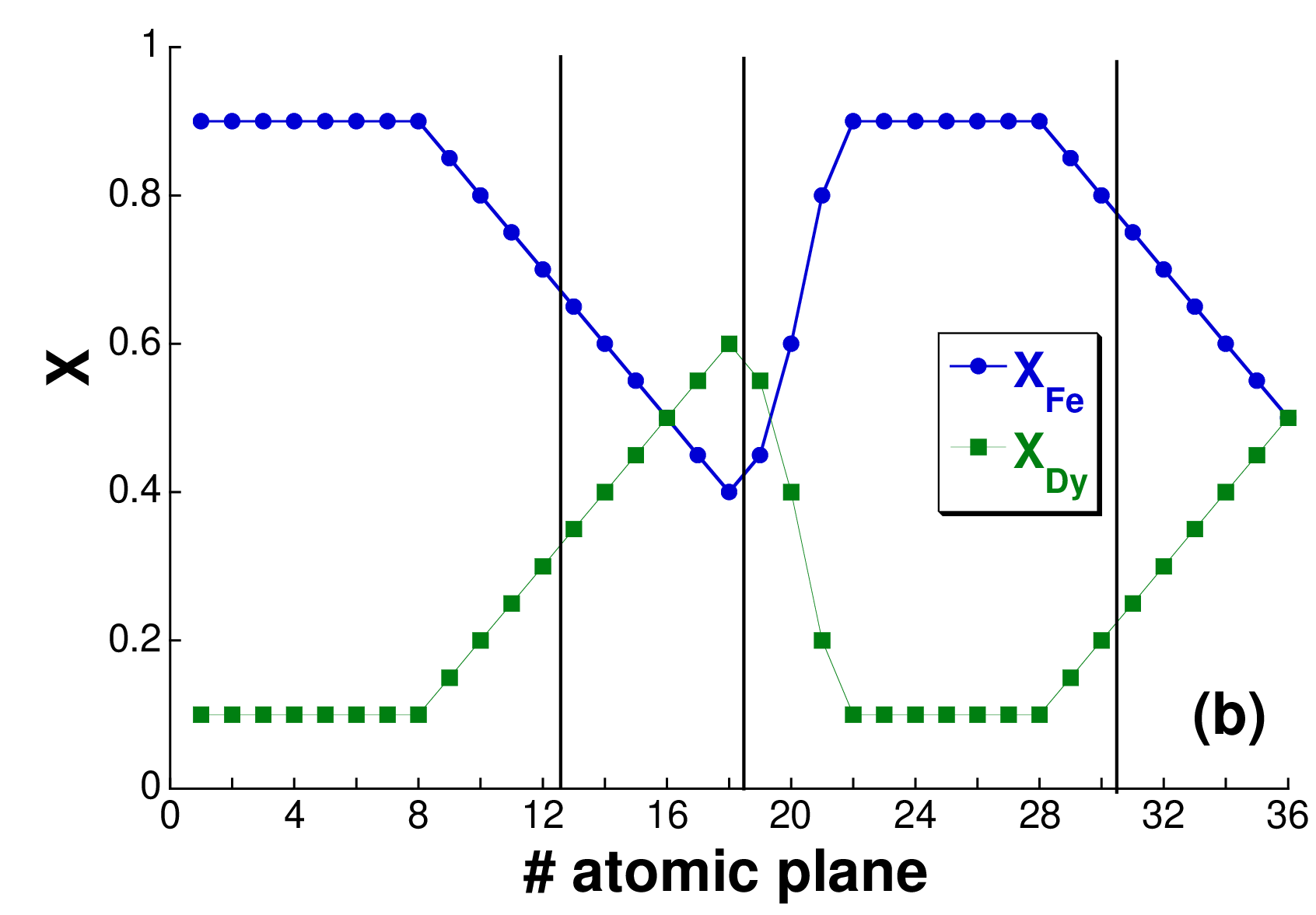}}
	\caption{Concentration profile of each specie along the perpendicular direction for an \textit{abrupt} (a) and a \textit{TAP} (b) concentration profile.}
\label{figure04}
\end{figure}
The introduction of realistic concentration profiles along the growth direction of the multilayer is essential since atomic diffusion seems to have a major influence on the macroscopic magnetic anisotropy. In this study, we have thus considered two concentration profiles: the first one, called \textit{abrupt}, corresponds to a multilayer made up of pure Fe layers and pure Dy layers with an abrupt interface (Figure \ref{figure04}(a)); the second profile (Figure \ref{figure04}(b)) is directly obtained from tomographic atom probe analyses of a multilayer (Fe 3nm/Dy 2nm) built up at 570K  \cite{tamion06} (Figure \ref{figure01}). This profile is composed of a rich Fe region (Fe$_{90}$Dy$_{10}$) and a wide region in which the concentration varies. As it has been previously mentioned, this concentration profile leads to the maximum of perpendicular magnetic anisotropy experimentally observed by Tamion et al. \cite{tamion06,tamion07}. In the following, this profile will be called \textit{TAP} profile (for tomographic atom probe profile). The main difference between the two profiles is that the pure Dy region in the multilayer with an abrupt profile is replaced by an alloyed region with variable local concentration in the case of the TAP profile.

\subsection{Magnetic anisotropy} \label{sect_clustermodel}

In order to investigate the perpendicular magnetic anisotropy in the Fe/Dy multilayers, we have focused our attention on a local structural anisotropy model \cite{mergel93} which seems to be well adapted to understand the referenced experimental results.

\begin{figure}\centering
{\includegraphics[width=0.35\textwidth]{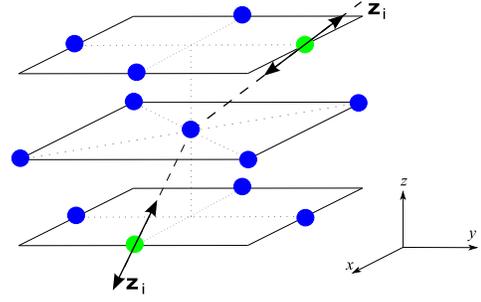}}
	\caption{Schematic representation of the ordered Fe-Dy clusters. Fe atoms are in blue, Dy atoms in green. The arrows show the local anisotropy axes on the Dy sites.}
\label{figure05}
\end{figure}

This model is based on the central hypothesis that during layer-by-layer growth small Fe-Dy crystallized zones at a scale of a few interatomic distances are formed defining on average a preferential axis perpendicular to the film plane. Indeed, they probably exhibit a structure that is reminiscent of those of the defined TM-RE compounds, whose hexagonal symmetry leads to anisotropy axes on average perpendicular to the layers. These locally ordered clusters consist in our model of 13 atoms (one central Fe atom and its 12 nearest neighbours) (Figure \ref{figure05}). Among the 8 neighbours which are not in the plane of the central atom, between 2 and 4 Dy atoms are randomly distributed in order to obtain a cluster concentration close to that of defined compounds. Since RE atoms are characterized by a strong intrinsic anisotropy due to the particular shape of their \textit{4f} orbitals, we have thus considered in the Hamiltonian a single site anisotropy term for all Dy atoms. The anisotropy axes on the Dy atoms belonging to the clusters are parallel to the Fe-Dy bonds (Fe here is the central atom of the cluster); so the unit vectors $\mathbf{z}_{i}$ of these 4 axes are $(\pm \frac{1}{\sqrt{2}}$,0,$\frac{1}{\sqrt{2}}$) or ($0,\pm \frac{1}{\sqrt{2}}$,$\frac{1}{\sqrt{2}}$). This leads on average to an anisotropy direction perpendicular to the plane of the layers. For the other Dy sites  which do not belong to the clusters we have considered random anisotropy because of the amorphous structure of the multilayers. 
The energy of the system is thus:
 
\begin{eqnarray}
\lefteqn{ E =  -\sum_{<i,j>} J_{ij} (\mathbf{S}_{i} . \mathbf{S}_{j}) - D_{\rm{Dy}} \bigg( \sum_{i \notin \rm{clusters}} (\mathbf{S}_{i}. \mathbf{n}_{i})^2 {} }
\nonumber\\
& & {}+\sum_{i \in \rm{clusters}} (\mathbf{S}_{i}.\mathbf{z}_{i})^2 \bigg),
\end{eqnarray} 

\noindent where ${\mathbf{S}_{i}}$ is the spin of the site $i$, $J_{ij}$ is the exchange interaction between the nearest-neighbour sites $i$ and $j$, $D_{{\rm Dy}}$ is the anisotropy constant on the Dy sites, ${\mathbf{n}_{i}}$ is an unit vector indicating the random anisotropy direction on the Dy site $i$ and ${\mathbf{z}_{i}}$ is an unit vector along the local anisotropy axis of a Dy site belonging to the clusters. We also define the magnetic susceptibility:
$$
	\chi (T) = \frac{\big< m^{2} \big>_{T} - \big< m \big>_{T}^{2}}{k_{\rm{B}}T} \times N,
	\label{eq:chi}
$$
where $\big< m \big>_{T}$ is the magnetisation defined as 
$$
	\big< m \big>_{T} = \frac{1}{N} \Bigg< \Bigg[ \bigg( \sum_{i=1}^{N} m_{i}^{x} \bigg)^{2} + \Big( \sum_{i=1}^{N} m_{i}^{y} \Big)^{2} + \Big( \sum_{i=1}^{N} m_{i}^{z} \Big)^{2} \Bigg]^{1/2} \Bigg>_{T}.
	\label{eq:aim}
$$

 As the anisotropy coefficient $D_{\mathrm{Dy}}$ in amorphous multilayers is not accurately determined in the literature, we will consider it as a free parameter with the same value for all the Dy sites (in the clusters and in the amorphous matrix). In the same way, the cluster concentration, which is defined as the number of atoms included in the clusters divided by the total number of atoms, is a free parameter in the simulations since it cannot be evaluated experimentally. But in this work, most of the results have been obtained with a quite small cluster concentration of 5\%.

\section{Numerical method} \label{sec3}

The numerical method is the importance sampling MC method in the canonical ensemble \cite{binder90,heermann90,mackeown97}. The thermodynamic equilibrium at each temperature, corresponding to the minimization of the free energy of the system, is obtained by the Metropolis algorithm \cite{metropolis53}. In this algorithm, the spins are examined individually. A site $i$ is randomly chosen and a unit vector $\mathbf{u}_{i}$ defined by the random choice with uniform distribution of its $z$-component $u_{i} \in [-1,1]$ and an its azimuthal angle $\varphi_{i} \in [0,2\pi[$ is determined. The new spin $\mathbf{S}^{'}_{i}$ is then given by:
$$
\big(S^{x}_{i}\big)^{'} = S_{i} \sqrt{1-u_{i}^{2}} \cos \varphi_{i},
$$
$$
\big(S^{y}_{i}\big)^{'} = S_{i} \sqrt{1-u_{i}^{2}} \sin \varphi_{i},
$$
$$
\big(S^{z}_{i}\big)^{'} = S_{i} u_{i},
$$
where $S_{i}=S_{\mathrm{Fe}}$ or $S_{\mathrm{Dy}}$. It has to be noted that this spin trial rotation procedure is isotropic. Then the energy variation $\Delta E$ associated to this rotation is calculated. The next step is the following:

\begin{itemize}
	\item if $\Delta E\leq 0$, the rotation is accepted;

	\item if $\Delta E > 0$, the rotation may be accepted with a probability that is proportional to the Boltzmann factor ${\rm exp}(-\Delta E/k_{\rm B} T)$ in order to take into account the thermal fluctuations.
\end{itemize}
	
	One MC step consists in examining all spins of the system once. At each temperature, 5000 MC steps were performed to reach the thermodynamic equilibrium, and afterwards the physical quantities were measured by averaging over the next 10000 MC steps. As we are interested in disordered systems, it is necessary to perform simulations on several disorder configurations (typically between 20 and 100) in order to estimate the thermodynamic quantities of a macroscopic system. It is important to note that for saving CPU time, we have limited our simulations to a double bilayer. In order to match up to the experimental system (Fe 3nm/Dy 2nm), we have considered 12 Fe and 6 Dy atomic planes in each bilayer because the ratio of the atomic Fe and Dy radii is around 1.5. Each plane contains 800 atoms (which gives a total number of atoms equal to 28800). Since we are not interested in the critical behaviour, finite size effects have no significant influence on the physical properties under consideration.

\section{Numerical results} \label{sec4}

We present here the magnetic properties of Fe/Dy multilayers with the concentration profiles defined previously (called \textit{abrupt} and \textit{TAP} profiles, see Figure \ref{figure04}). As a first step, since we are interested in the influence of the concentration profile on the magnetic ordering, we have considered a multilayer without any anisotropy term. The next step of this study is then focused on the effect of the magnetic anisotropy on the Dy atoms in the clusters and in the amorphous matrix.

\subsection{Influence of the concentration profile without anisotropy ($D_{\mathrm{Dy}}=0$)}
\begin{figure}\centering
\rotatebox{-90}{\includegraphics[width=0.3\textwidth]{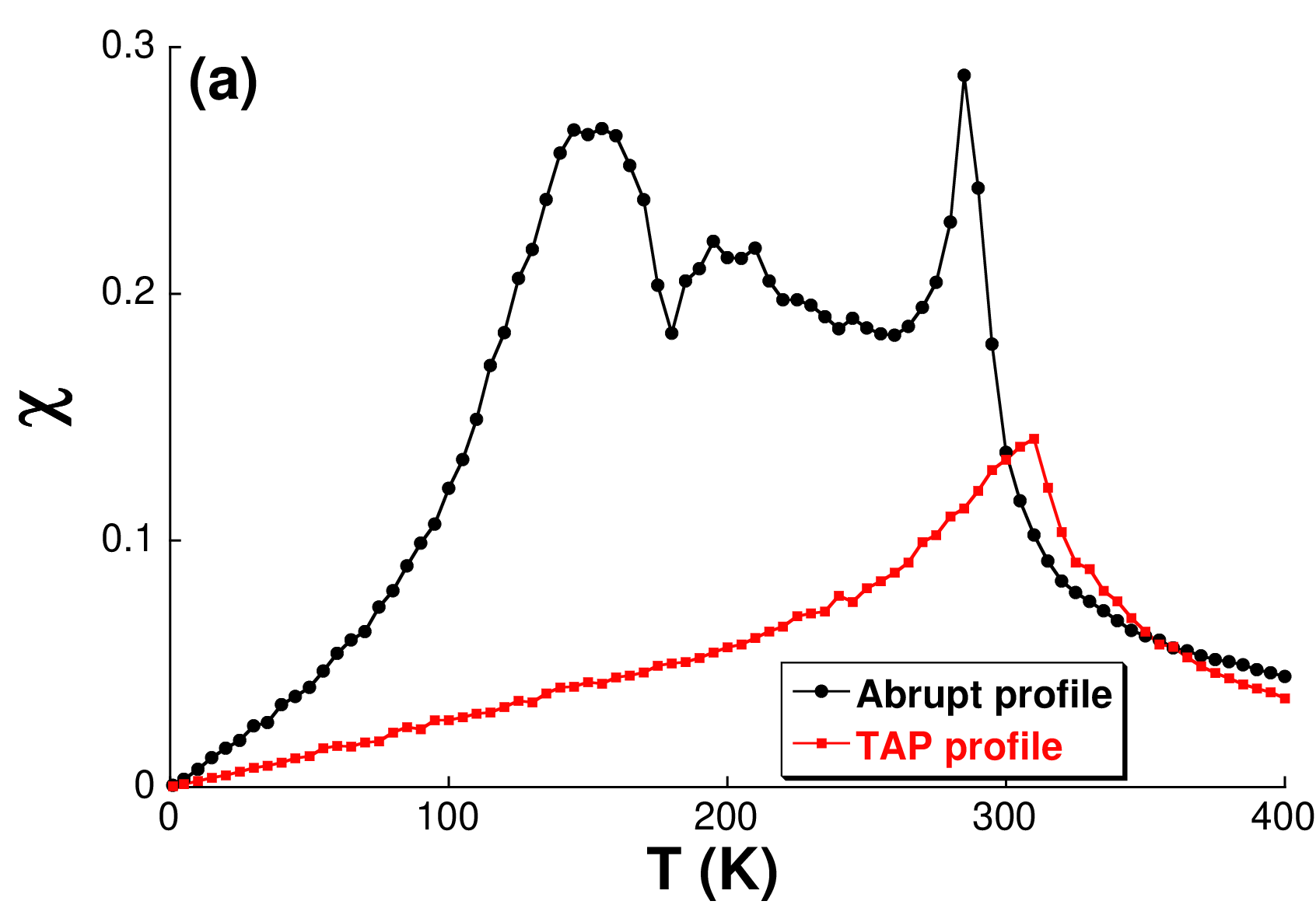}}
\rotatebox{-90}{\includegraphics[width=0.3\textwidth]{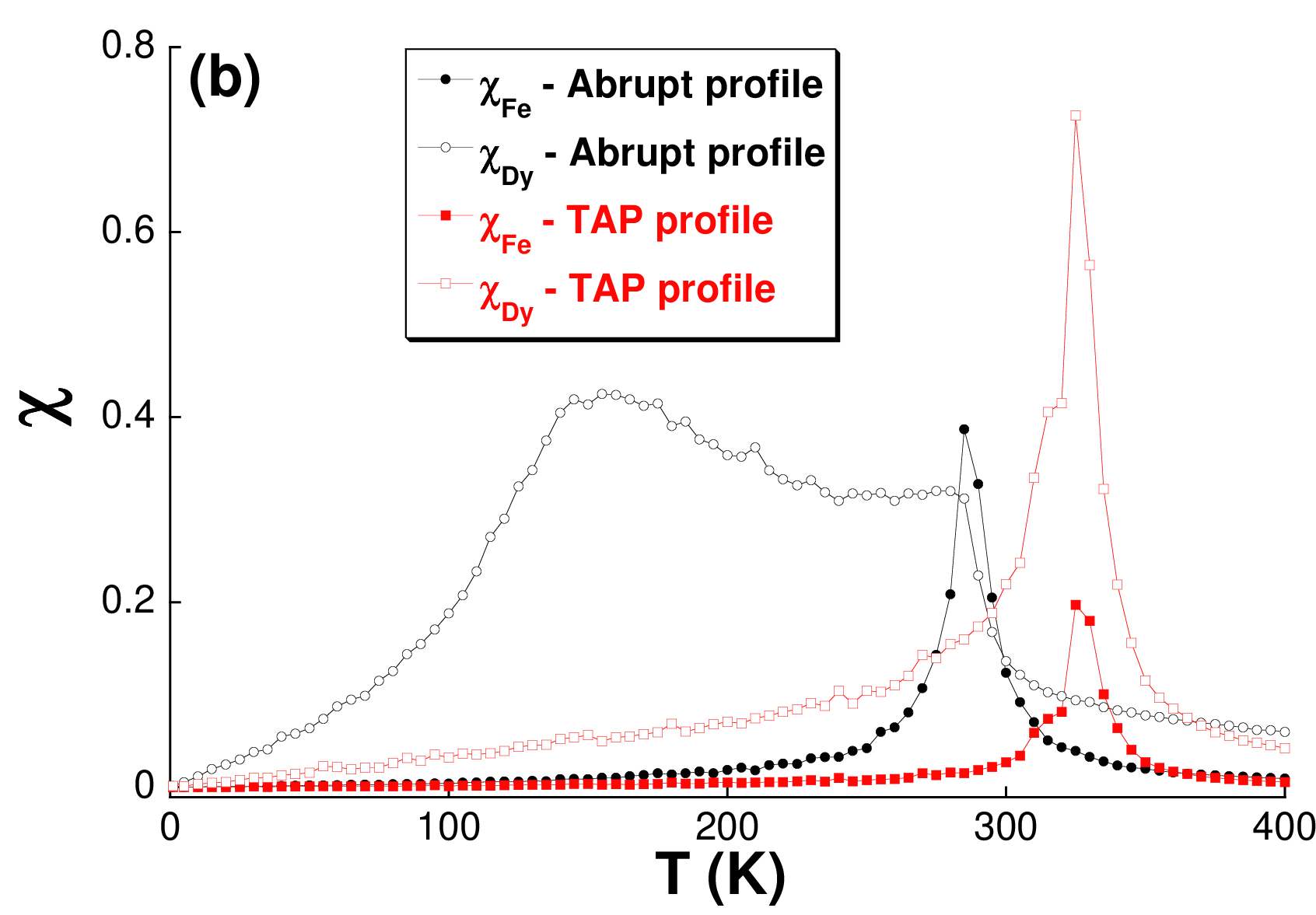}}
	\caption{Thermal variation of the magnetic susceptibility for a multilayer (Fe 3nm/Dy 2nm) with
 the 2 studied concentration profiles ((a) for the whole sample, (b) per sublattice). The curves for the \textit{TAP} profile result from the average over 4 disorder configurations.}
\label{figure06}
\end{figure}
\begin{figure}\centering
\rotatebox{-90}{\includegraphics[width=0.3\textwidth]{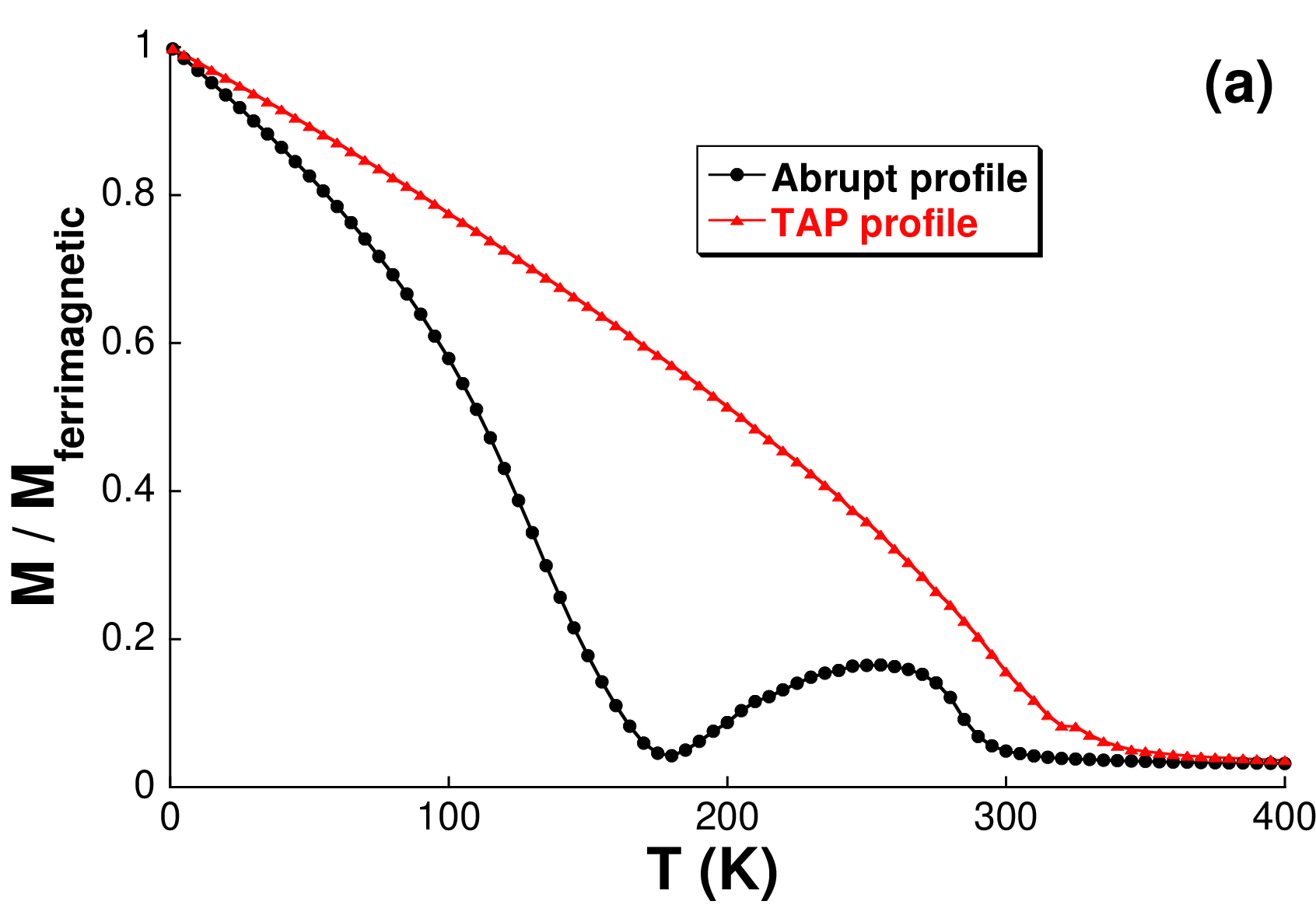}}
\rotatebox{-90}{\includegraphics[width=0.3\textwidth]{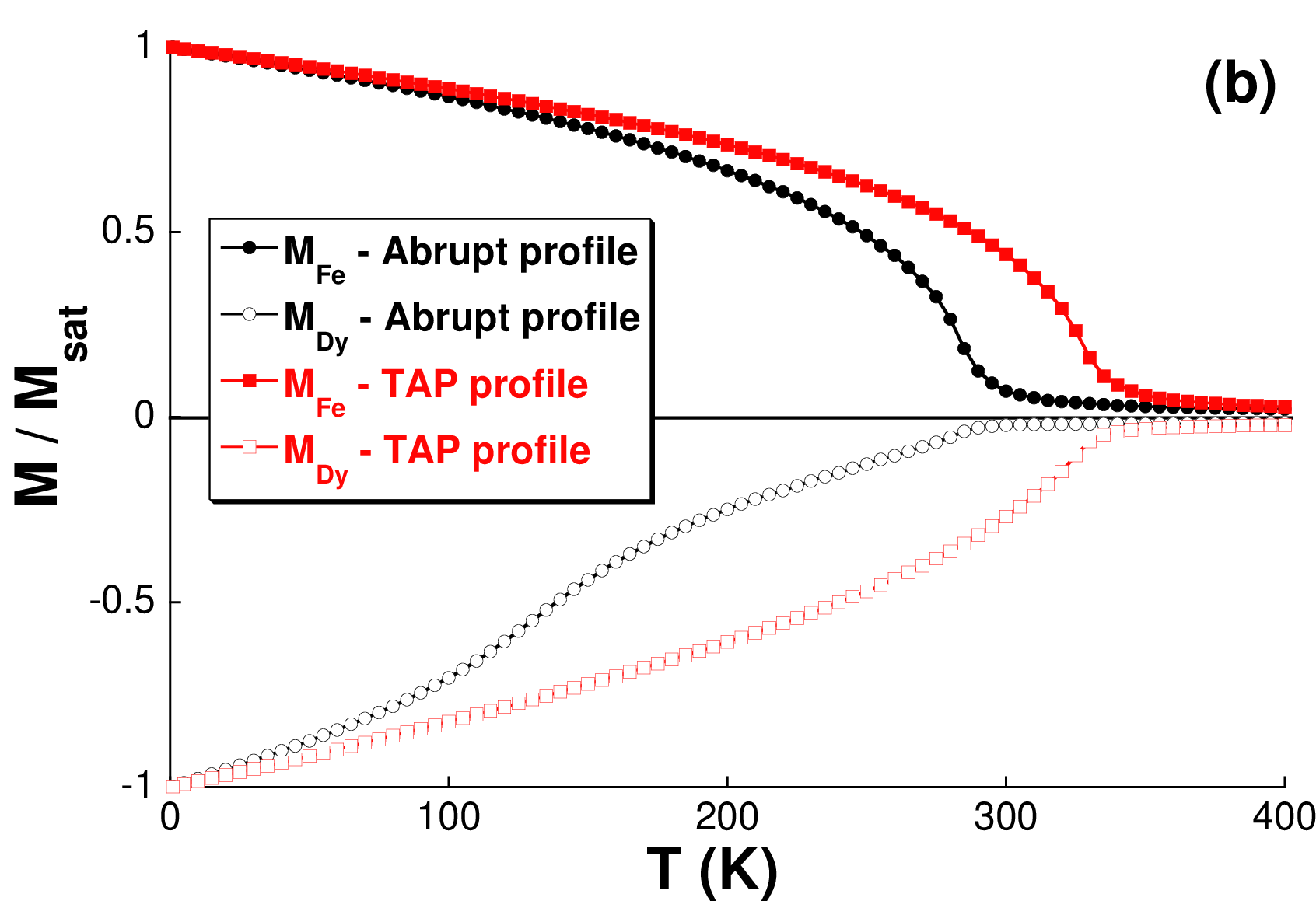}}
	\caption{Thermal variation of the reduced magnetisation for a multilayer (Fe 3nm/Dy 2nm) with the 2 concentration profiles ((a) for the whole sample, (b) per sublattice). $M_{{\rm ferrimagnetic}}$ (a) corresponds to the ground states with antiparallel Fe and Dy layer magnetisations whereas $M_{{\rm sat}}$ (b) is the saturated magnetisation of each ferromagnetic sublattice.}
\label{figure07}
\end{figure}

No anisotropy term has been introduced into the Hamiltonian, this latter is then invariant under any rotation of all the spins. In these samples, the ground states consist of a ferromagnetic Fe layer and a ferromagnetic Dy layer which are antiparallel. These ground states, so called ferrimagnetic, will be considered as a reference in the following. The thermal variation of the magnetic susceptibility for the whole sample and for each sublattice is shown in Figure \ref{figure06} with the \textit{abrupt} and \textit{TAP} concentration profiles. We observe that the \textit{abrupt} profile is characterized by two separate magnetic ordering at respectively $285$K and $150$K corresponding to the Fe layer and the Dy layer. The Dy susceptibility curve (Figure \ref{figure06}(b)) exhibits a shoulder at $T \sim 280$K related to the polarisation, via the exchange interaction, of the Dy atoms located in the neighbourhood of the interface. For the \textit{TAP} profile, the two sublattices order at the same temperature ($325$K). The shoulder of the Dy susceptibility curve  below the peak temperature corresponds to the more concentrated Dy planes. The thermal variation of the magnetisation of the whole Fe/Dy multilayer (Figure \ref{figure07}(a)) is strongly influenced by the concentration profile even if the concentration of each specie is the same in the two cases ($X_{\mathrm{Fe}}=2/3$).

 We note in the case of the \textit{abrupt} profile the presence of a compensation point at $T=175$K when the two sublattice magnetisations are antiparallel with the same modulus. Concerning the \textit{TAP} profile, the magnetic ordering is continuous and smooth because of the mixing of the two species.

\subsection{Influence of the magnetic anisotropy: sperimagnetic order}

\subsubsection{Metastable states}

The ground states result from the competition between exchange interactions and magnetic anisotropy. The exchange interactions lead to a collinear order in these systems since there is no frustration whereas the magnetic anisotropy favours a angular distribution of the magnetic moments. Then the ground states are expected to be sperimagnetic with a cone angle that should increase with the anisotropy constant $D_{\mathrm{Dy}}$. 

As it has been described in the section \ref{sect_clustermodel}, due to the locally ordered clusters, the average orientation of the moments should be perpendicular to the layers. Because of the competition between exchange and anisotropy energies, numerical convergence towards metastable states at low temperature can occur during the simulation, mainly in the case of small cluster concentrations and small values of $D_{\rm Dy}$.

\begin{figure}\centering
\rotatebox{-90}{\includegraphics[width=0.3\textwidth]{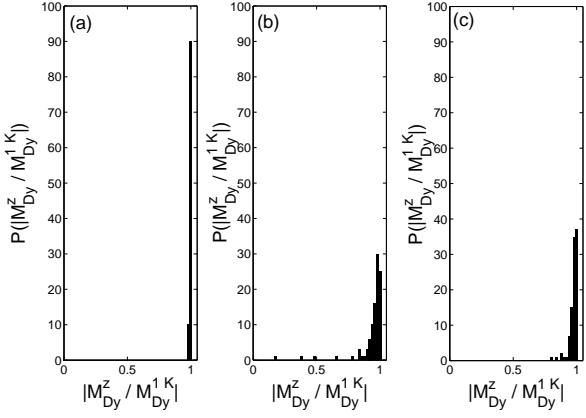}
}
\caption{Histograms over 100 disorder configurations at $T=1$K of the perpendicular magnetisation component of the Dy sublattice measured with the \textit{TAP} profile, a concentration of clusters of $10\%$  ((a) without any RMA and $D_{\rm Dy}/k_{\rm{B}}=10$K, (b) with RMA on the matrix Dy atoms and $D_{\rm Dy}/k_{\rm{B}}=10$K, (c) with RMA on the matrix Dy atoms and $D_{\rm Dy}/k_{\rm{B}}=50$K).}
  \label{figure08}
\end{figure}

Figure \ref{figure08} shows the histograms over 100 chemical disorder configurations of the perpendicular magnetisation component of the Dy sublattice for a multilayer with the \textit{TAP} profile  at $T=1$K. The concentration of ordered clusters is of $10\%$ and $D_{\rm Dy}/k_{\rm{B}}=10$K. Without any  random magnetic anisotropy (RMA) on the Dy atoms of the matrix (Figure \ref{figure08}(a)), we observe a sharp distribution at $\vert M_{\mathrm{Dy}}^{z} / M_{\mathrm{Dy}}  \vert \sim 1$ which means that the magnetisation of the Dy sublattice is perpendicular to the layers for all disorder configurations. The introduction of RMA on the Dy atoms of the matrix leads to a broader distribution $P(\vert M_{\mathrm{Dy}}^{z} / M_{\mathrm{Dy}} \vert)$ (Figure \ref{figure08}(b)). In particular, some disorder configurations exhibit a magnetisation which is clearly non perpendicular to the layers. This phenomenon is less pronounced when the anisotropy constant $D_{\mathrm{Dy}}$ increases (Figure \ref{figure08}(c)). Anyway, this problem has been overcome, by decreasing the cooling rate during the simulation and also increasing the number of MC steps at each temperature. 

\subsubsection{Influence of the concentration profile for different $D_{\mathrm{Dy}}$ values}
Here we have studied the influence of the magnetic anisotropy constant on the Dy sites in a multilayer containing $5\%$ of locally ordered clusters which means that only $3.5\%$ of Dy atoms exhibit uniaxial anisotropy (on average).

$\bullet$ \underline{Abrupt profile}

\begin{figure}\centering
{\includegraphics[angle=-90,width=0.4\textwidth]{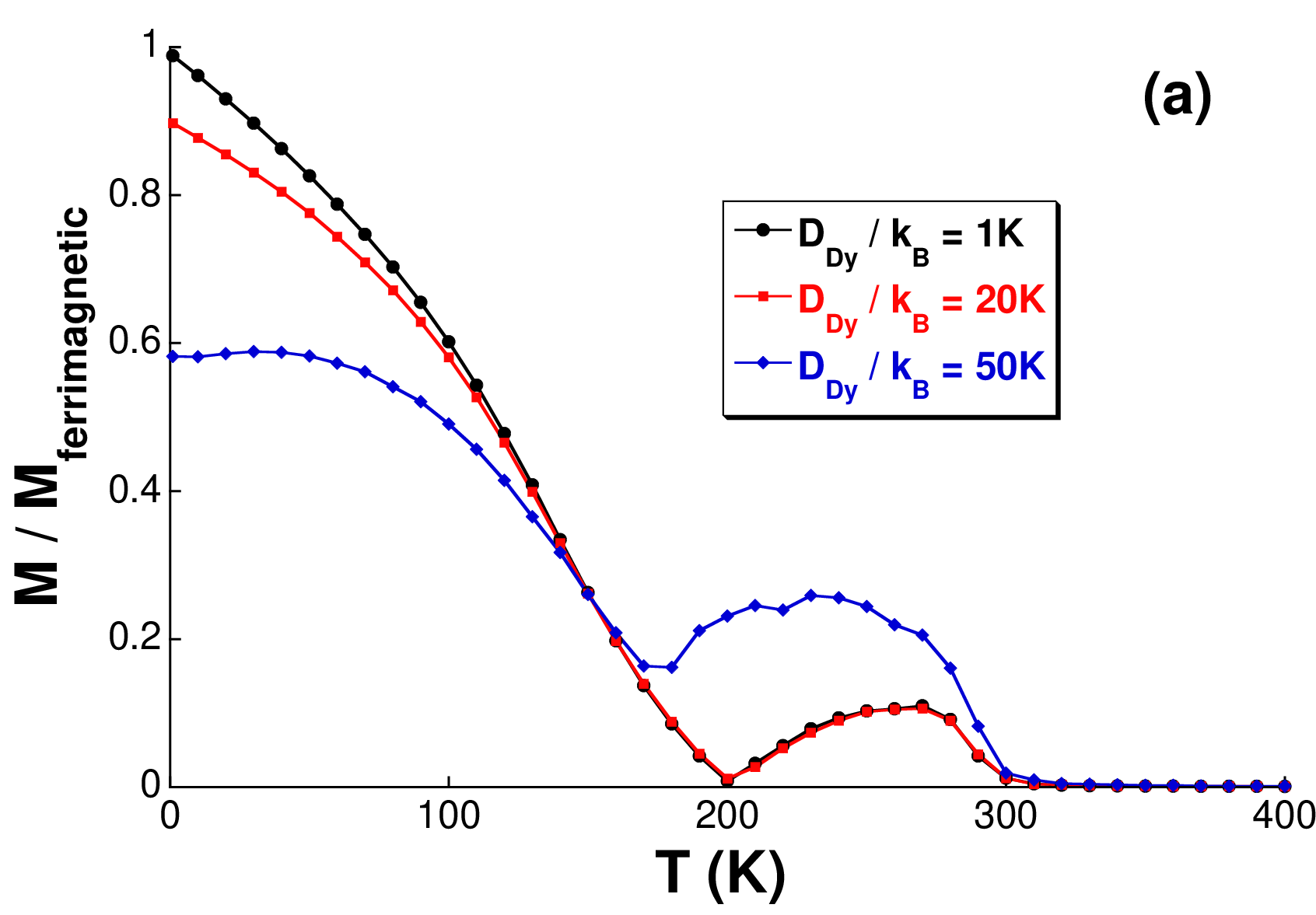}
\includegraphics[angle=-90,width=0.4\textwidth]{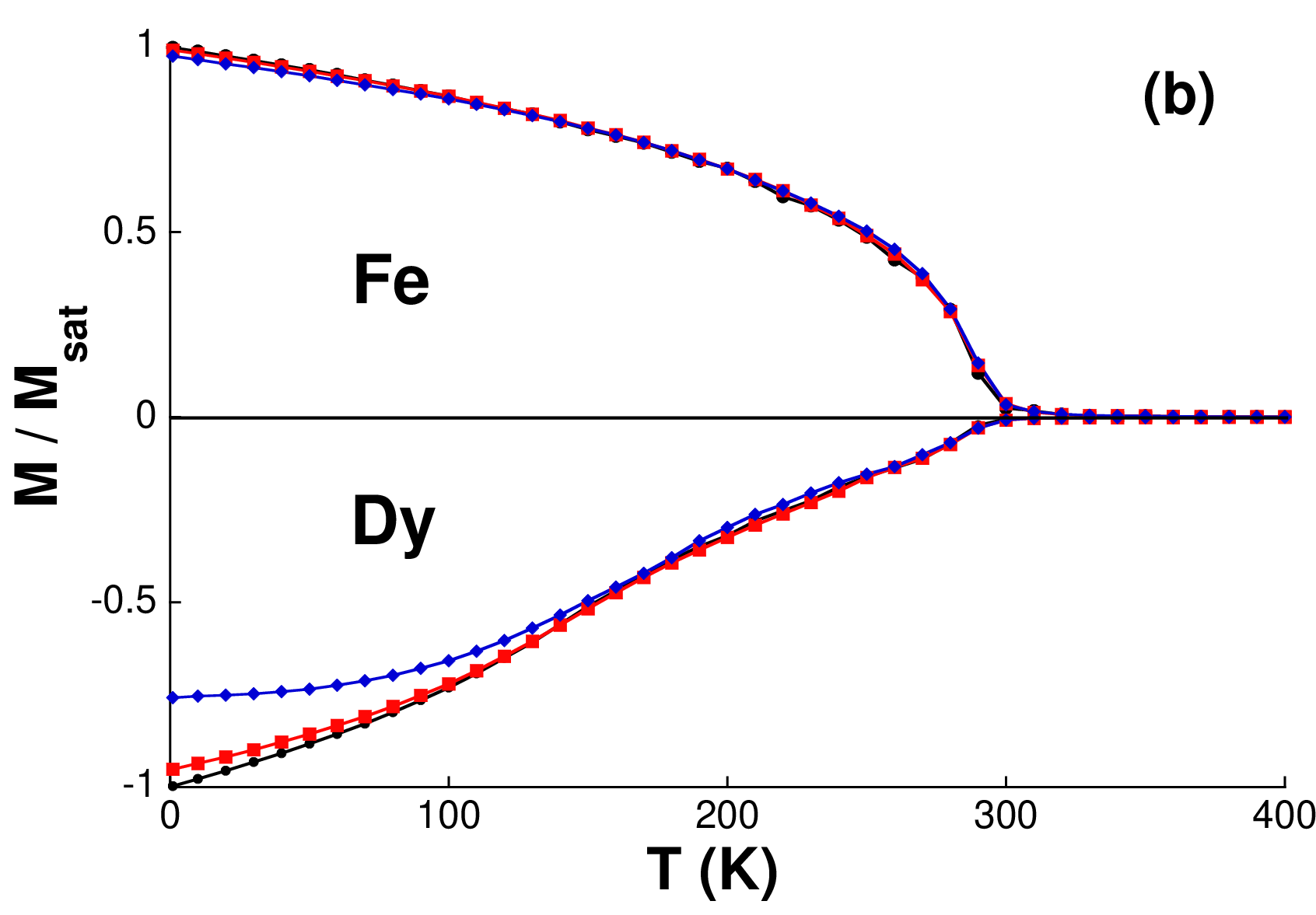}}
	\caption{Thermal variation of the reduced magnetisation for a multilayer (Fe 3nm/Dy 2nm) with an \textit{abrupt} concentration profile containing 5$\%$ of clusters for different values of the Dy anisotropy constant $D_{\rm Dy}$ ((a) total magnetisation, (b) Fe and Dy sublattice magnetisations).}
    \label{figure09}
\end{figure}
\begin{figure}\centering
\rotatebox{-90}{\includegraphics[width=0.33\textwidth]{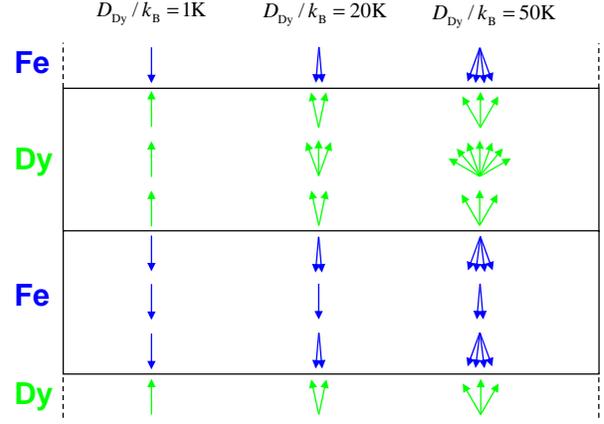}}
	\caption{Schematic representation of the magnetic order at low temperature in a multilayer with an \textit{abrupt} concentration profile for 3 different anisotropy constant values $D_{\rm Dy}$.}
	\label{figure10}
\end{figure}

We firstly presente the results concerning the \textit{abrupt} concentration profile. Figure \ref{figure09} shows the thermal variation of the global magnetisation and of the Fe and Dy sublattices for different values of $D_{\rm Dy}$. We observe a strong influence of $D_{\rm Dy}$ on the low temperature magnetisation ($T<T_{\rm comp}=175$K) (Figure \ref{figure09}(a)), especially on the Dy sublattice, its influence on the Fe sublattice being imperceptible (Figure \ref{figure09}(b)). This phenomenon can be understood by the fact that the magnetic ordering process above $T_{\rm comp}$ is essentially due to the Fe magnetic moments which do not display any magnetic anisotropy. Below $T_{\rm comp}$, both the low-temperature global magnetisation and the Dy sublattice magnetisation decrease when the anisotropy constant increases. This is due to the increase of the non collinearity of the Dy moments, that is a sperimagnetic order already described in the literature in similar systems \cite{coey78,hansen89}. These magnetic configurations are qualitatively shown in Figure \ref{figure10} for three values of $D_{\rm{Dy}}$. The angular distribution of Dy moment is broader in the core of the Dy layer whereas it is the contrary for the Fe moments which are more misaligned close to the interface because of Fe-Dy coupling.

\begin{figure}\centering
{\includegraphics[angle=-90,width=0.4\textwidth]{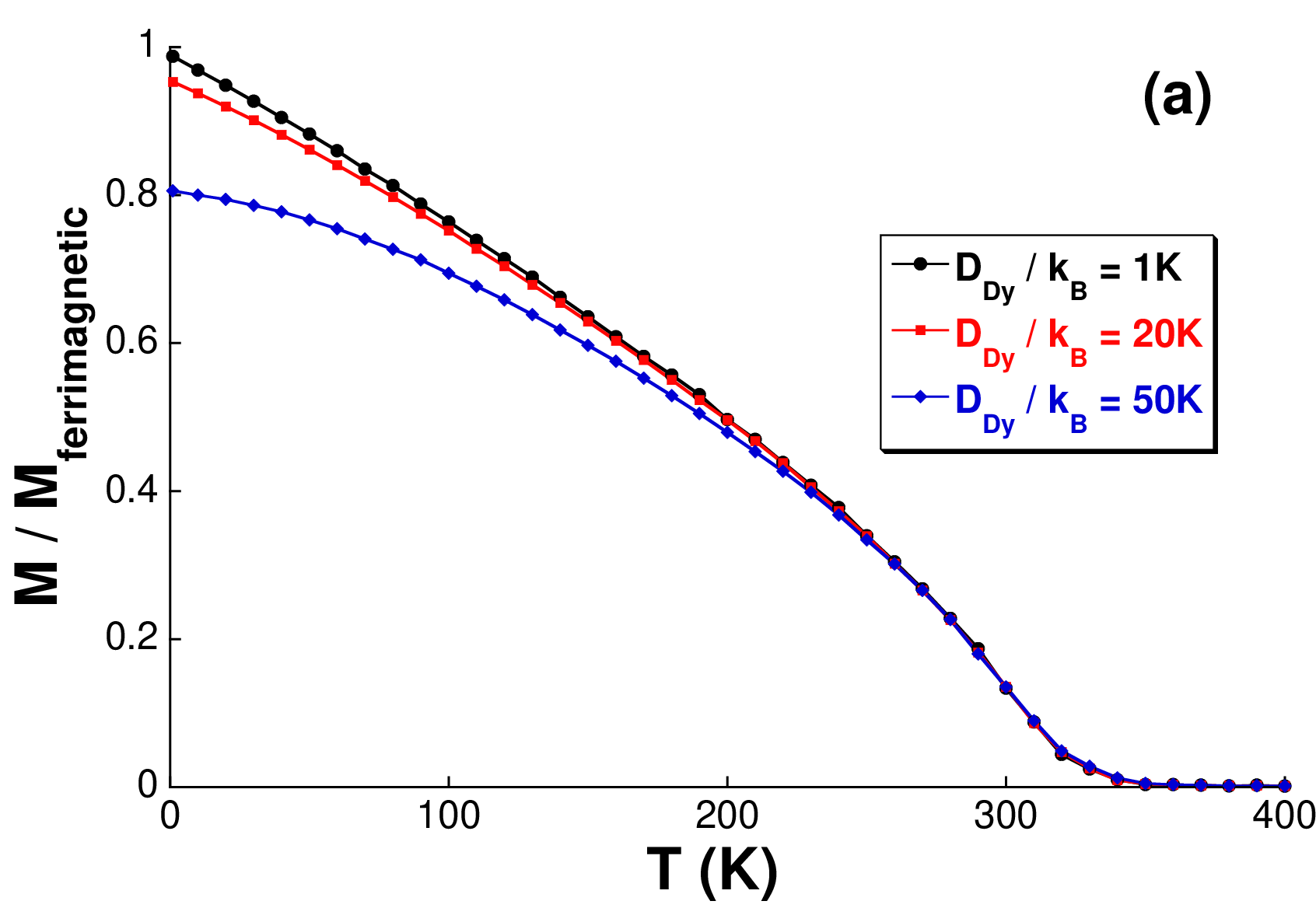}
\includegraphics[angle=-90,width=0.4\textwidth]{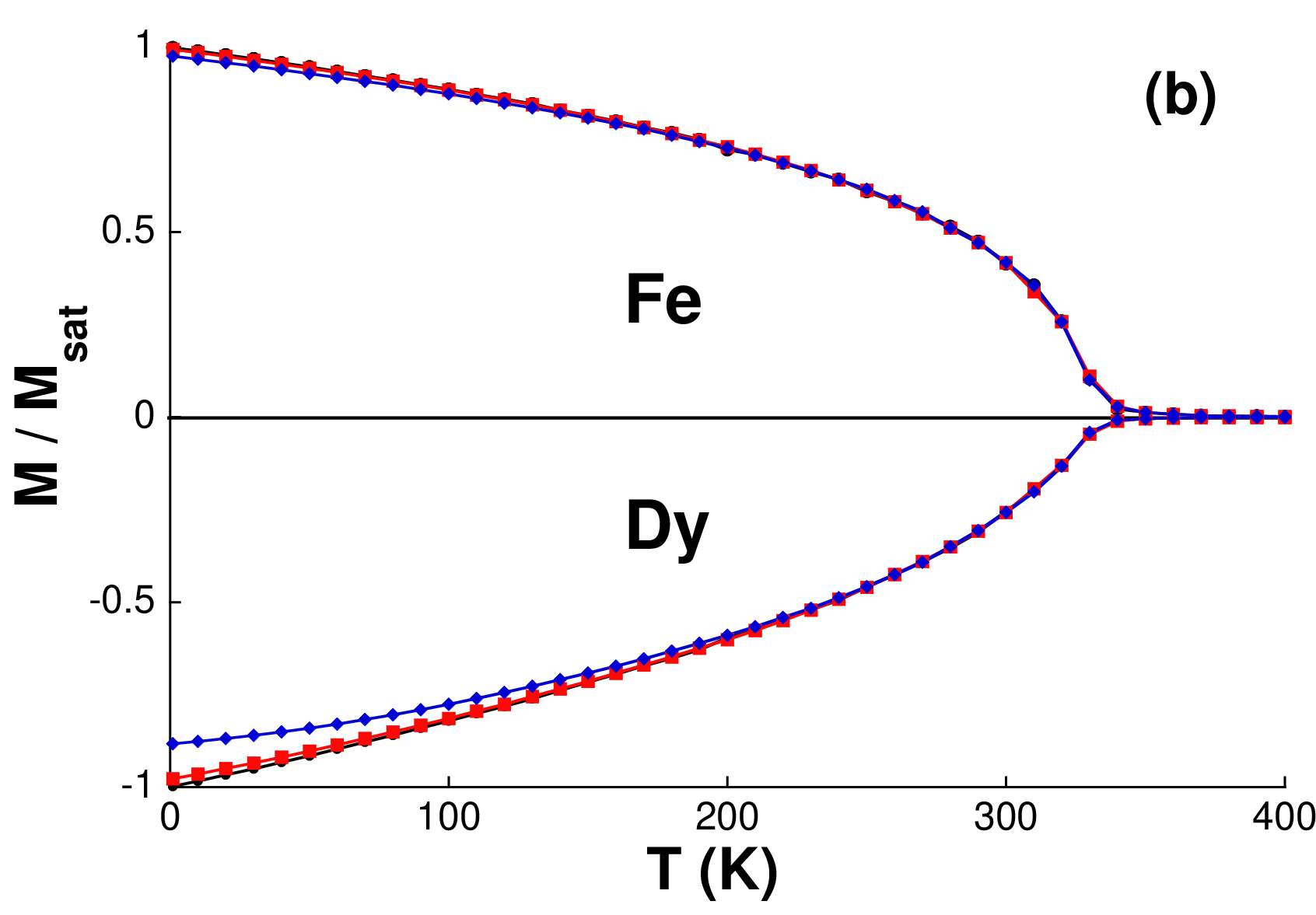}}
	\caption{Thermal variation of the reduced magnetisation for a multilayer (Fe 3nm/Dy 2nm) with a \textit{TAP} concentration profile containing 5$\%$ of clusters for different values of the Dy anisotropy constant $D_{\rm Dy}$ ((a) total magnetisation, (b) Fe and Dy sublattice magnetisations).}
    \label{figure11}
\end{figure}

$\bullet$ \underline{TAP profile}\\
The influence of the concentration profile on the sperimagnetic order has been investigated, using the previously defined \textit{TAP} profile. Figure \ref{figure11} shows the thermal variation of the global magnetisation and of the Fe and Dy sublattices for a concentration of ordered clusters of $5\%$ and different values of $D_{\rm{Dy}}$.

\begin{figure}\centering
\rotatebox{-90}{\includegraphics[width=0.33\textwidth]{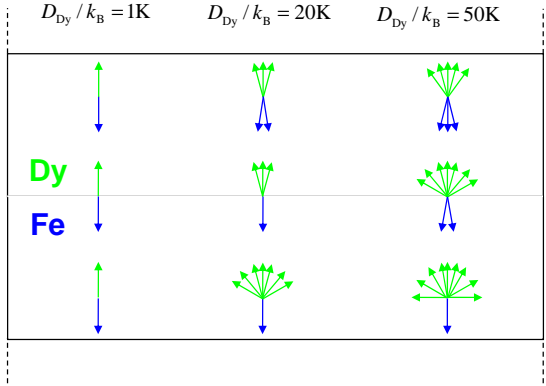}}
	\caption{Schematic representation of the magnetic order at low temperature in a multilayer with a \textit{TAP} concentration profile for 3 different anisotropy constant values $D_{\rm Dy}$.}
	\label{figure12}
\end{figure}

As in the case of the \textit{abrupt} profile, we observe a decrease of the global magnetisation of the multilayer when $D_{\rm{Dy}}$ increases but this decrease is much less marked than in the previous case. Indeed, we note that the magnetisation of the Dy sublattice is less influenced by $D_{\rm{Dy}}$ whereas a small effect on the Fe sublattice can be seen when $D_{\rm{Dy}}/k_{\rm{B}}=50$K. It means that as the layers are more mixed, the influence of the RMA on the Dy atoms is partially transmitted to the Fe sublattice which displays also a low temperature non-collinear magnetic order (Figure \ref{figure12}). We can see that, at least for $D_{\rm{Dy}}/k_{\rm{B}}=50$K, the angular distribution of the Fe moments broadens while those of the Dy moments sharpens.
\subsubsection{Influence of the concentration profile on the magnetisation profiles ($D_{\mathrm{Dy}}/k_{\mathrm{B}}=50$K)}

\newpage
\begin{figure*}
\centering
\rotatebox{-90}{\includegraphics[width=0.57\textwidth]{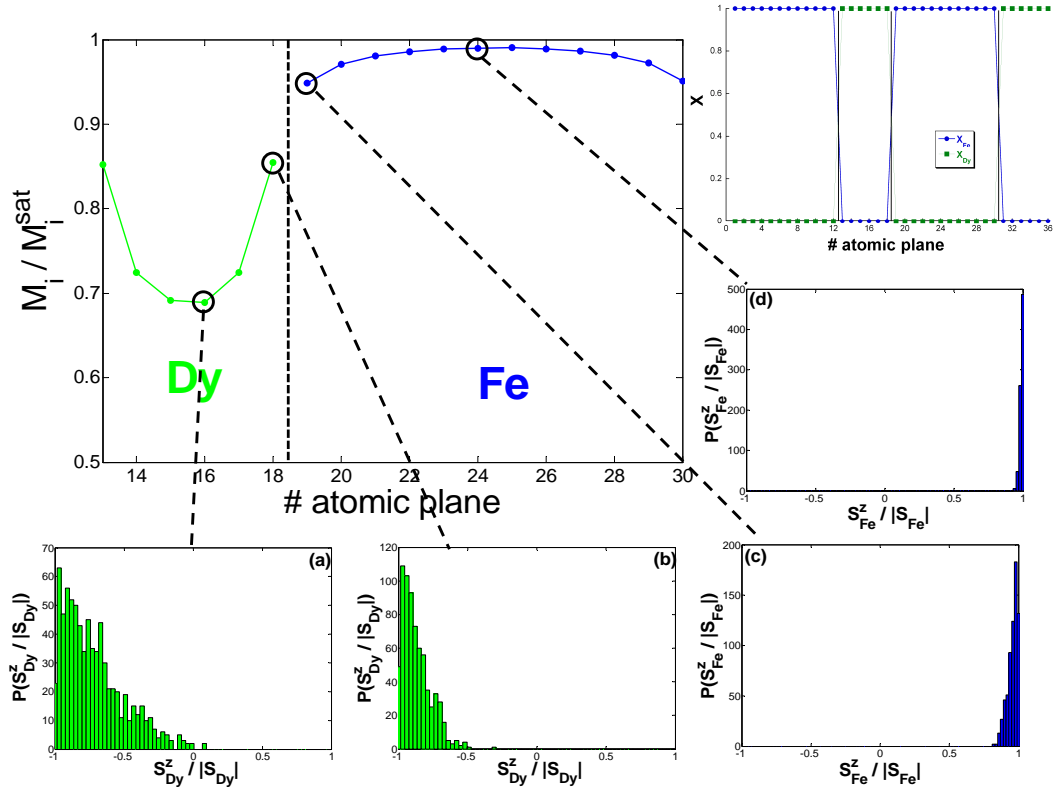}}
\caption{Reduced magnetisation of each atomic plane of a multilayer at low temperature ($T=1$K) with the \textit{abrupt} concentration profile containing $5\%$ of clusters and $D_{\rm{Dy}}/k_{\rm{B}}=50$K. Distribution of the perpendicular magnetic component in the Dy layer and in the Fe layer ((a) Dy - core plane; (b) Dy - interface plane; (c) Fe - interface plane; (d) Fe - core plane).}
\label{figure13}
\end{figure*}
\begin{figure*}
\centering
\rotatebox{-90}{\includegraphics[width=0.57\textwidth]{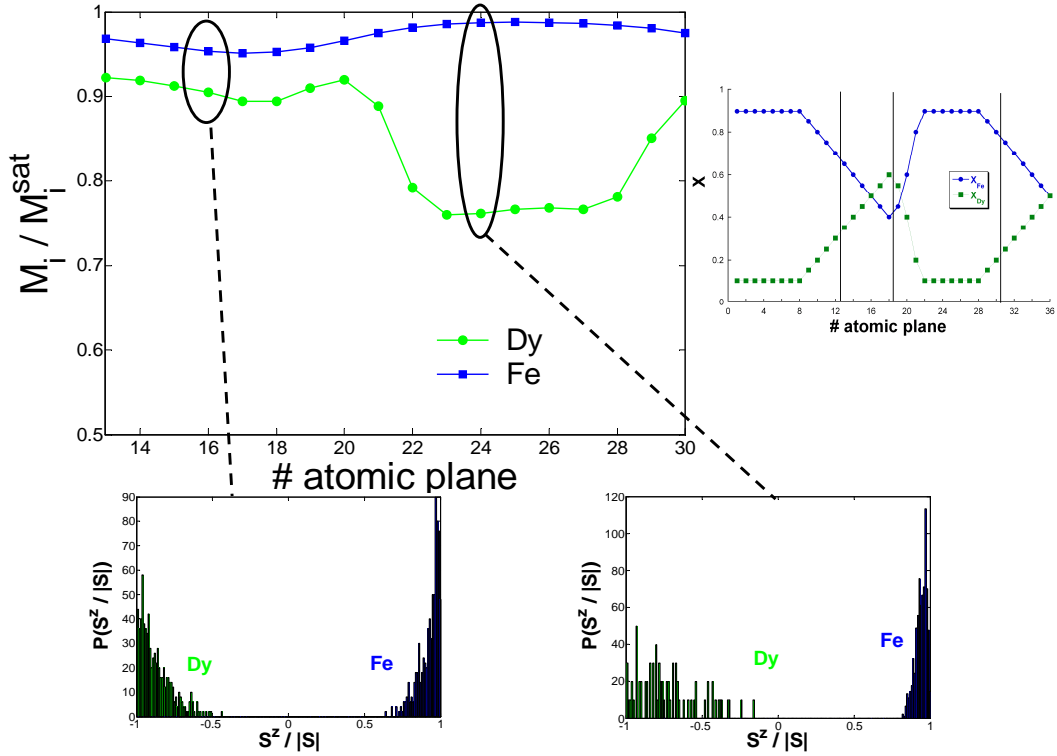}}
	\caption{Reduced magnetisation of each atomic plane of a multilayer at low temperature ($T=1$K) with the \textit{TAP} concentration profile containing $5\%$ of clusters and $D_{\rm{Dy}}/k_{\rm{B}}=50$K. Distribution of the perpendicular magnetic moment component in the Dy sublattice and in the Fe sublattice ((a) plane 16 - $X_{\mathrm{Fe}}=0.50$; (b) plane 24 - $X_{\mathrm{Fe}}=0.90$) }
\label{figure14}
\end{figure*}
\newpage

 In order to get a quantitative analysis of the sperimagnetism, we have calculated the magnetisation for each plane and the histogram of the perpendicular component of the magnetic moments for each plane and each sublattice. Since the sperimagnetism is more pronounced for $D_{\rm{Dy}}/k_{\rm{B}}=50$K, we have chosen this case. The results obtained at $T=1$K are shown in Figures \ref{figure13} and \ref{figure14} for the \textit{abrupt} and \textit{TAP} profiles respectively.

For the \textit{abrupt} profile, we clearly see the evolution of the magnetisation per plane which is not homogeneous inside the multilayer (Figure \ref{figure13}). The decrease of the magnetisation is related to the broadening of the distribution of the perpendicular component of the atomic moments as mentioned previously. More precisely, far from the interface, the distribution is very broad in the Dy atomic planes (Figure \ref{figure13}(a)) due to the large value of the ratio  $D_{\mathrm{Dy}}/J_{\mathrm{Dy-Dy}}=7.7$. At the interface, the distribution width is minimum in the Dy layer (Figure \ref{figure13}(b)) whereas it is maximum in the Fe layer (Figure \ref{figure13}(c)) because of the strong Fe-Dy coupling. Finally, the distribution is close to a Dirac peak in the core of the Fe layer since there is no magnetic anisotropy (Figure \ref{figure13}(d)).

In the case of the \textit{TAP} profile (Figure \ref{figure14}), the variation of the magnetisation per plane for each sublattice is different from the previous case since it depends on the local concentration. We observe that the magnetisation of the Fe sublattice reaches a maximum for the high-concentrated Fe planes ($X_{\mathrm{Fe}}=0.9$) whereas for the Dy sublattice, the magnetisation per plane exhibits a maximum for a plane with $X_{\mathrm{Fe}}$ close to $0.4$. This agrees with the distribution of the perpendicular component of the magnetic moments for the Dy sublattice which is significantly sharper in planes with  $X_{\mathrm{Fe}}=0.4$. Contrarily, for the Fe sublattice, the distribution is sharper in the rich Fe planes. In order to explain these features, we have reported in Table \ref{Tab_1}, the magnitude of each energy term (per atom) defined as:
$$E_{\mathrm{RMA}}^{\mathrm{Dy}} = (1-X_{\mathrm{Fe}}-X_{\mathrm{Dy}}^{\mathrm{cluster}} ) D_{\mathrm{Dy}} S_{\mathrm{Dy}}^{2},$$
$$E_{\mathrm{exch}}^{\mathrm{Dy-Dy}} = (1-X_{\mathrm{Fe}})^{2} J_{\mathrm{Dy-Dy}} S_{{\mathrm{Dy}}}^{2},$$
$$E_{\mathrm{exch}}^{\mathrm{Fe-Dy}} = X_{\mathrm{Fe}}(1-X_{\mathrm{Fe}}) J_{\mathrm{Fe-Dy}} S_{\mathrm{Fe}}S_{\mathrm{Dy}},$$
$$E_{\mathrm{exch}}^{\mathrm{Fe-Fe}} = X_{\mathrm{Fe}}^{2}  J_{\mathrm{Fe-Fe}} S_{\mathrm{Fe}}^{2},$$
where $X_{\mathrm{Dy}}^{\mathrm{cluster}}$ is the Dy atomic fraction in the clusters. In the 24$^{\mathrm{th}}$ plane (rich-Fe plane), the coupling between the Dy and Fe sublattices is weak, then the Fe moments are almost collinear whereas the Dy moments are strongly out of line due to the large ratio ${E_{\mathrm{RMA}}^{\mathrm{Dy}}}/{E_{\mathrm{exch}}^{\mathrm{Dy-Dy}}+E_{\mathrm{exch}}^{\mathrm{Fe-Dy}}} \sim 5.2$. Concerning the 16$^{\mathrm{th}}$ and 13$^{\mathrm{th}}$ atomic planes, the Dy-Fe sublattice coupling is strong and the possible sperimagnetism of the Dy sublattice (due to the very large $E_{\mathrm{RMA}}^{\mathrm{Dy}}$ value) could be transmitted to the Fe sublattice. The main difference between these two planes is the ratio $E_{\mathrm{RMA}}^{\mathrm{Dy}}/(E_{\mathrm{exch}}^{\mathrm{Dy-Dy}} + E_{\mathrm{exch}}^{\mathrm{Fe-Dy}}) $ which is about 1.9 and 2.1 in the 13$^{\mathrm{th}}$ and 16$^{\mathrm{th}}$ atomic planes respectively. This explains why the magnetic configuration is more collinear in the 13$^{\mathrm{th}}$ plane than in the 16$^{\mathrm{th}}$ plane. The comparison of these two values with those of the 24$^{\mathrm{th}}$ plane accounts for the more marked sperimagnetism of the Dy sublattice in this latter.

\begin{table}[!ht]
\caption{\label{Tab_1}Random magnetic energy and exchange interaction energy for 3 typical atomic planes for the \textit{TAP} profile.}
\begin{tabular}{cccccc}
\hline
plane &$X_{\mathrm{Fe}}$ &$E_{\mathrm{RMA}}^{\mathrm{Dy}}$ (K) & $E_{\mathrm{exch}}^{\mathrm{Dy-Dy}}$ (K) & $E_{\mathrm{exch}}^{\mathrm{Fe-Dy}}$ (K) & $E_{\mathrm{exch}}^{\mathrm{Fe-Fe}}$ (K) \\
\hline
13 &0.65&  102.34&  4.97 & 49.25 & 49.34 \\
16 &0.50 & 146.88 & 10.16 & 61.14 & 21.77 \\
24 &0.90 & 29.30 & 0.41 & 5.24 & 86.29 \\
\hline
\end{tabular}
\end{table}

These results evidence that the magnetisation profile along the multilayer is not homogeneous, which has already been observed experimentally on these samples from polarized neutrons reflectivity measurements \cite{tamion07}. We would like to emphasize that unlike what is mentionned in several papers \cite{coey76,gu90,ishio93}, our results indicate that the atomic moment distribution around the mean magnetisation direction in each plane is not uniform but rather gaussian for the two concentration profiles.

 \subsubsection{Influence of the cluster concentration}
The influence of the cluster concentration also has been investigated for $D_{\mathrm{Dy}}/k_{\mathrm{B}}=50$K in the case of the \textit{TAP} profile. 

\begin{figure}[!ht]
\centering
{\includegraphics[angle=-90,width=0.4\textwidth]{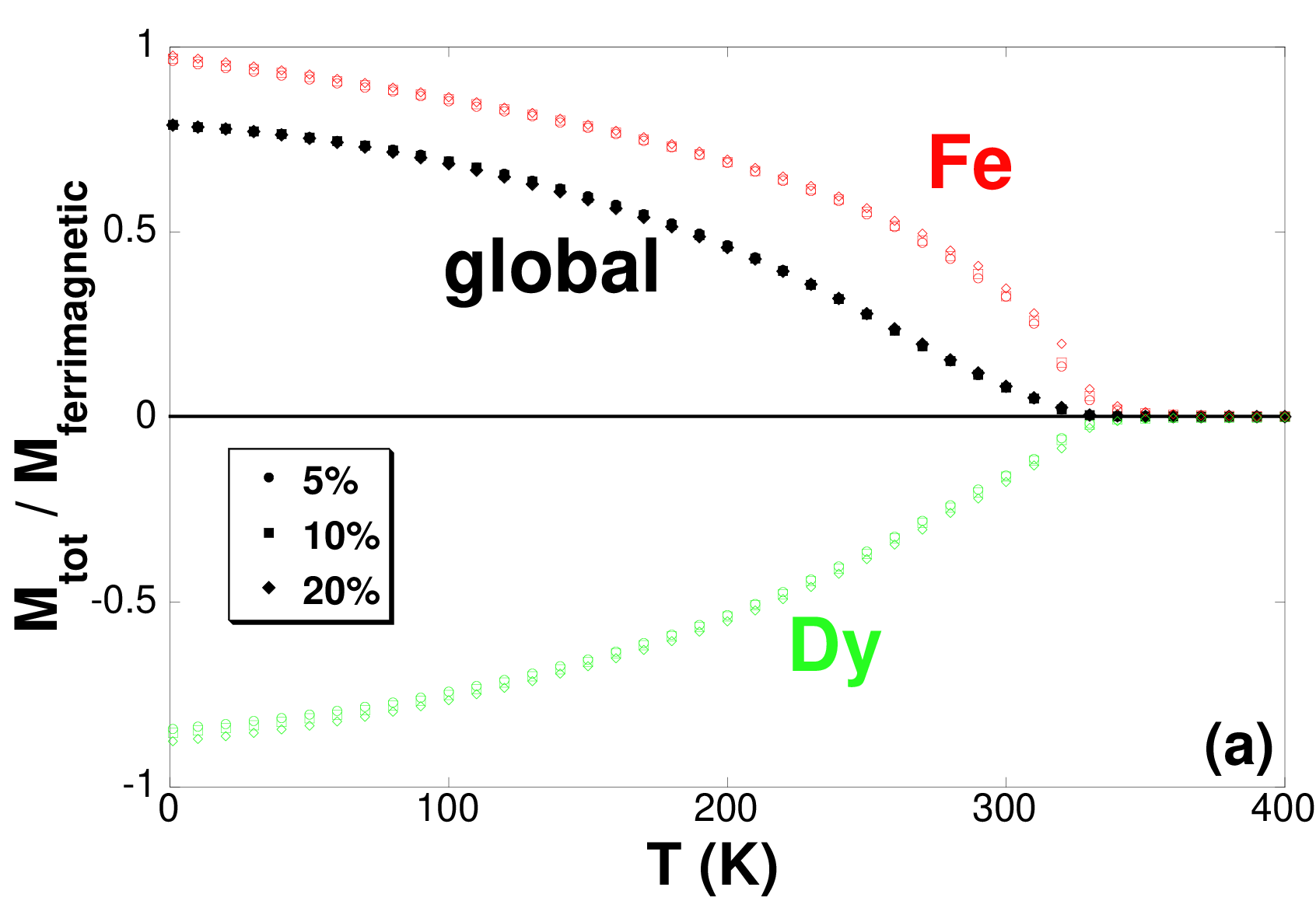}
\includegraphics[angle=-90,width=0.4\textwidth]{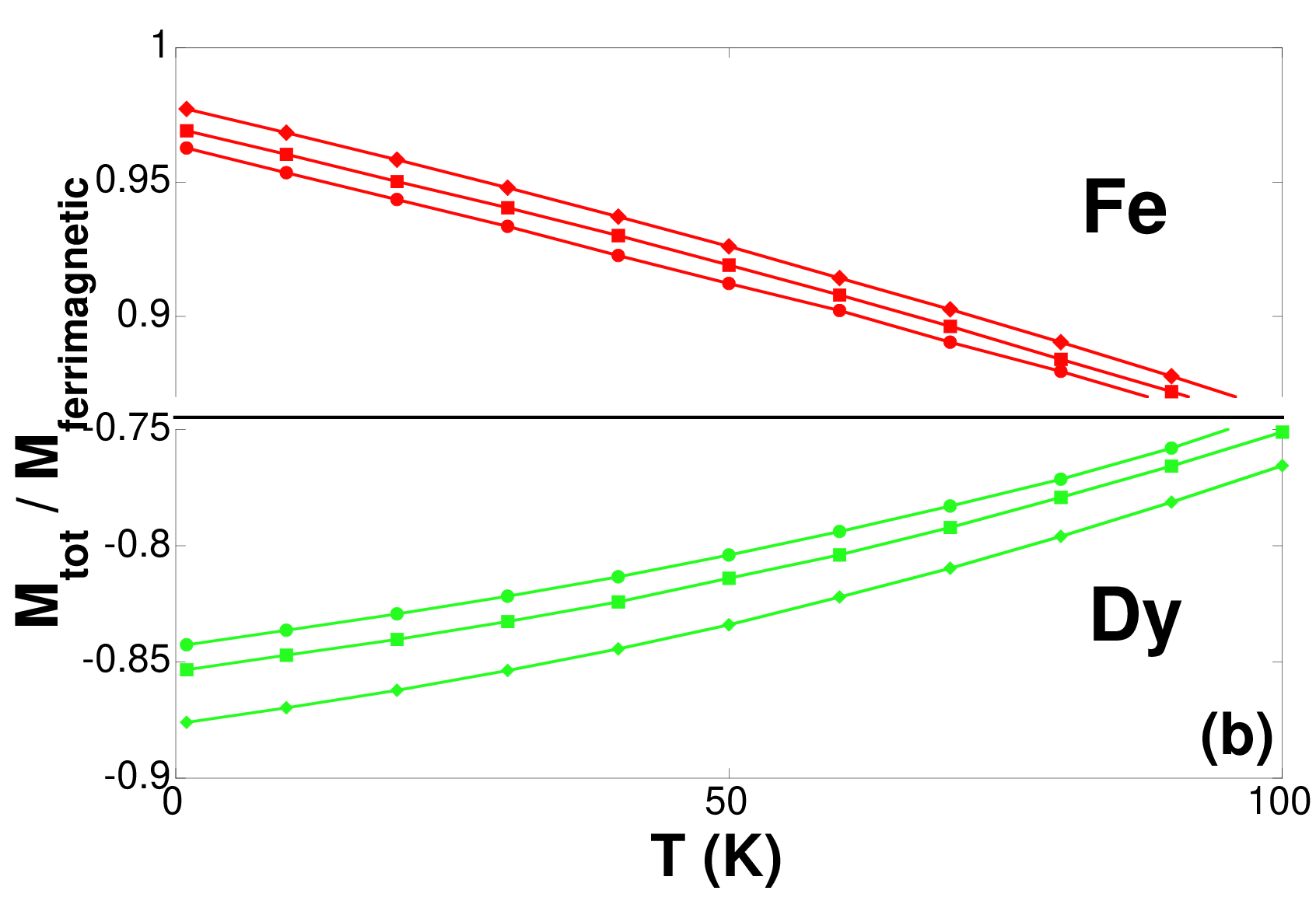}}
	\caption{Thermal variation of the reduced magnetisation for a multilayer (Fe 3nm/Dy 2nm) with a \textit{TAP} concentration profile for $D_{\mathrm{Dy}}/k_{\mathrm{B}}=50$K for different cluster concentrations ((a) total and sublattice magnetisation, (b) magnification of the sublattice magnetisation).}
\label{figure15}
\end{figure}

Figure \ref{figure15} shows the thermal variation of the reduced magnetisation for such a multilayer for different cluster concentrations. The increase of the cluster concentration from 5\% to 20\% has no influence on the global magnetisation: the three curves for each concentration are superimposed at each temperature (Figure \ref{figure15}(a)). Looking more precisely at the sublattice magnetisations at low temperature (Figure \ref{figure15}(b)), we note a very small increase of each sublattice magnetisation as the cluster concentration rises. Quite surprisingly, the cluster concentration has no significant effect on the angular distribution of the moments unlike the anisotropy constant to which the magnetic order is very sensitive. This result can be interpreted from the strong influence of the exchange interactions in these multilayers which induce a global behaviour of the magnetic moments as previously seen.

\section{Conclusion} \label{sec5}
 In order to qualitatively describe the macroscopic perpendicular anisotropy, we have used in our model a local structural anisotropy model which induces a perpendicular magnetisation even for very small cluster concentrations. This study has shown using a numerical investigation the influence of the concentration profile on the thermal variation of the magnetisation and on the magnetisation profile for several values of $D_{\mathrm{Dy}}$. For the \textit{abrupt} profile, a compensation point can be observed and sperimagnetism occurs essentially in the Dy layer. Concerning the \textit{TAP} profile, the sperimagnetism on the Dy sublattice is not homogeneous and is less pronounced in the rich Dy plane with $X_{\mathrm{Dy}} \sim 0.60$. In a near future, we plan to perform simulations of hysteresis loops in order to explain experimental data by suggesting rotation mecanisms for the magnetic moments in function of the concentration profile, the anisotropy constant $D_{\mathrm{Dy}}$ and the cluster concentration.

\ack
The simulations were performed at the Centre de Ressources Informatiques de Haute Normandie (CRIHAN) under the project No. 2004002. Moreover, the authors are indebted to Dr. Alexandre Tamion for valuable discussions on the experimental results.



\begin{thebibliography}{00}
\bibitem{sato86} N. Sato, {J. Appl. Phys.} {59} (1986) 2514. 
\bibitem{shan90} Z. S. Shan and D. J. Sellmyer, {Phys. Rev.} B {42} (1990) 10433.
\bibitem{schulte95} O. Schulte, F. Klose and W. Felsh, {Phys. Rev.} B {52} (1995) 6480.
\bibitem{mergel93} D. Mergel, H. Heitmann and P. Hansen, {Phys. Rev.} B {47} (1993) 882.
\bibitem{fuji96} H. Fujiwara, X. Y. Yu, S. Tsunashima, S. Iwata, M. Sakurai and K. Suzuki, {J. Appl. Phys.} {79} (1996) 6270.
\bibitem{shan90b} Z. S. Shan, D. J. Sellmyer, S. Jaswal, Y. Wang and J. Shen, {Phys. Rev.} B {42} (1990) 10446.
\bibitem{tamion05} A. Tamion, E. Cadel, C. Bordel and D. Blavette, {J. Magn. Magn. Mater.} {290-291} (2005) 238.
\bibitem{tamion06} A. Tamion, E. Cadel, C. Bordel and D. Blavette, {Scripta Mater.} {54} (2006) 671.
\bibitem{tamion07} A. Tamion, F. Ott, P. E. Berche, E. Talbot, C. Bordel and D. Blavette, to be published in {J. Magn. Magn. Mater.} (2008).
\bibitem{heiman76} N. Heiman, K. Lee, R. I. Potter and S. Kirkpatrick, {J. Appl. Phys.} {47} (1976) 2634.
\bibitem{mimura78} Y. Mimura, N. Imamura, T. Kobayashi, A. Okada and Y. Kushiro, {J. Appl. Phys.} {49} (1978) 1208.
\bibitem{hansen91} P. Hansen, S. Klahn, C. Clausen, G. Much and K. Witter, {J. Appl. Phys.} {69} (1991) 3194.
\bibitem{kobe76} S. Kobe and K. Handrich, {Phys. Stat. Sol. (b)} {73} (1976) K65.
\bibitem{tappert96} J. Tappert, W. Keune, R. A. Brand, P. Vulliet, J.-P. Sanchez and T. Shinjo, {J. Appl. Phys.} {80} (1996) 4503.
\bibitem{binder90} K. Binder and D. W. Heermann, {Monte Carlo Simulation in Statistical Physics}, Springer, Berlin (1997).
\bibitem{heermann90} D. W. Heermann, {Computer Simulation Methods in Theoretical Physics}, Springer, Berlin (1990).
\bibitem{mackeown97} P. K. Mac Keown, {Stochastic Simulation in Physics}, Springer, Berlin (1997).
\bibitem{metropolis53} N. Metropolis, A. Rosenbluth, M. Rosenbluth, A. Teller and E. Teller, {J. Chem. Phys.} {21} (1953) 1087.
\bibitem{coey78} J. M. D. Coey, {J. Appl. Phys.} {49} (1978) 1646.
\bibitem{hansen89} P. Hansen, C. Clausen, G. Much, M. Rosenkranz and K. Witter, {J. Appl. Phys.} {66} (1989) 756.
\bibitem{coey76} J. M. D. Coey, J. Chappert, J. P. Rebouillat and T. S. Wang, {Phys. Rev. Lett.} {36} (1976) 1061.
\bibitem{gu90} B. X. Gu, H. R. Zhai and B. G. Shen, {Phys. Rev. B} {42} (1990) 10648.
\bibitem{ishio93} S. Ishio, N. Obara, S. Negami, T. Miyazaki, T. Kamimori, H. Tange and M. Goto, {J. Magn. Magn. Mater.} {119} (1993) 271.
\end{thebibliography}
\end{document}